\documentstyle{ article}

\begin{document}

\bigskip

\begin{center}
{ \Large \bf On Homological and
Homotopical Algebra of  Supersymmetries and Integrability in String Theory
   }
\end{center}
\smallskip
\begin{center}
{\bf Nikolaj M. Glazunov } \end{center}

\smallskip
\begin{center}

{\rm National  Aviation University.\\
     Prospect Komarova 1
            03058, Kiev-58 GSP Ukraine } \\
{\it glanm@yahoo.com }

 \end{center} \smallskip

 \begin{center} {\bf Abstract} \end{center}

The text contains introduction and preliminary definitions and results to my talk on
category theory description of  supersymmetries
and integrability in string theory. In the talk I plan to present homological and
homotopical algebra framework for Calabi-Yau supermanifolds and stacks in open and
closed string theory. In the framework  we investigate supersymmetries and integrability.

\section{Introduction}
In this paper we plan to discuss homological and homotopical algebra description of  supersymmetries
and integrability in string theory.
Recall that the Picard groupoid on an open set $U$ is the category whose objects are line bundles on $U$ and whose
morphisms are isomorphisms.  A gerbe is a stack over a topological space which is locally isomorphic to the Picard
groupoid of the space. A description of a stack or a gerbe in terms of an atlas with relations is known as a
presentation. Every stack has a presentation of the form of a global quotient stack $[X/G]$, for some space $X$ and
some group $G.$
However, presentations are not unique. A given stack can have many different presentations of the form $[X/G].$
Review very shortly some selected works in the directions.
Authors of the paper\cite{PS} investigate gauged linear sigma model descriptions of toric  stacks.
The paper includes (i) the gauged linear sigma model (GLSM)
description of toric stacks; (ii) a describing of the physics of GLSM; (iii) checking that physical predictions of
those GLSM exactly match the corresponding stacks. The description of Deligne-Mumford stacks
over toric varieties (toric stacks) is given with the help of stacky fans by authors of \cite{BCS}.
  It is well known that the GLSM is closely related to toric geometry. Some ground facts can be found in the
book~\cite{HKKPTVVZ}.
In the absence of a superpotential, the set of supersymmetric ground states of the GLSM is a toric variety.
Conversely, toric varieties can be described as the set of ground states of an appropriate gauged linear sigma model.
String theories on orbifolds were intoduced by Dixon, Harvey,  Vafa, and  Witten in \cite{DHVW}.
Recently very interesting results were obtained in the local theory of current-algebraic orbifolds\cite{HH}.\\
 Some integrable models are connected with solitons. Bogomol'nyi vortices are static, topologically stable, finite energy
 solution of the critically coupled Abelian Higgs model in $2 + 1$ dimensions~\cite{JT,MN}.
  Monopoles are solitons in three dimension.
  The Yang-Mills equations are differential equations which are hyperbolic in Minkowski
space-time, but have elliptic  counterparts when Minkowski space-time is replaced by Euclidian space. In both
(hyperbolic and elliptic) cases soliton type solutions (magnetic monopoles and instantons respectively)
have been extensively studied~\cite{Do,UY}.
A class of $D-$branes for the type $IIB$ plane-wave background as well as classical
description of the $D-$ instanton in the light-cone gauge  can be found in \cite{GG}.\\
There is a correspondence between rational conformal
field theories  of $SU(2)-$type~\cite{Zu}
and double triangle algebra associated to an ADE graph~\cite{CT,Tr}.\\
 After the introductory section and preliminary results  we plan to consider
next topics in the framework of Calabi-Yau supermanifolds and stacks in open
and closed string theory:

\begin{itemize}
\item Homological algbra of connections on manifolds and variation of Hodge structure
\item Homotopical categories
\item PROPs and Operads
\item $A_{\infty}-$categories
\item $A_{\infty}-$functors
\end{itemize}

\section{Examples, Defintions and Previous results}

\subsection{Superalgebras and their representations}

One of the simplest example of superalgebra\cite{VS} is connected with
supersymmetric
oscillator. Let $n_{B}, n_{F}$ be bosonic and fermionic infill numbers. State vectors of the
system are defined by

$$ | n_{B}, n_{F} \rangle, n_{B} = 0, 1, 2, \ldots, \infty;  n_{F} = 0, 1.$$

Let $b^{+}, b^{-}, f^{+}, f^{+}$  be  creation  and annihilation  bosonic and fermionic operators
respectively. Operators act on state vectors $ | n_{B}, n_{F}\rangle $ and vary infill numbers
$n_{B}, n_{F}$ by standard manner~\cite{GK}.
 Operators  $b^{+}, b^{-}, f^{+}, f^{+}$ satisfy following  commutator and anticommutator relations:
$$ [b^{+}, b^{-}] = 1,   { f^{+}, f^{-}} = 1, (f^{+})^{2} = (f^{-})^{2}= 0, [b,f] = 0. $$

Let $Q_{+}, Q_{-}$ operators that transform a boson to a fermion and vise versa. In previous  notations
$$ Q_{+} = q b^{-} f^{+},  Q_{-} = q b^{+} f^{-},$$
where $q$ is an arbitrary constant. Hermitian operators
$$ Q_{1} = Q_{+} + Q_{-}, Q_{2} = i(Q_{+} + Q_{-}),$$
satisfy anticommutator relation
$$ \{Q_{1}, Q_{2}\} = 0.  $$

The supersymmetric Hamiltonian is defined by
$$ H = Q_{1}^{2} = Q_{2}^{2}= \{ Q_{+}, Q_{-}\},$$
 It satisfies relation
$$[H,Q] = 0,$$
where $Q$ is one of $Q_{+},  Q_{-}, Q_{1}, Q_{2}.$ The Lie superalgebra is defined by
relations $$\{Q_{l}, Q_{k}\} =  2\delta_{lk}H, l, k = 1,2, $$
$$ [Q_{k}, H] = 0 .$$

\subsection{Categories of algebras}
Let $k$ be a field, ${\mathcal  A}lg_{k}$ the category of algebras over $k.$
 Objects of ${\mathcal  A}lg_{k}$ are $k-$algebras and morphisms
 are $k-$homomorphisms of the algebras. ${\mathcal  A}lg_{k}$
 contains many subcategories. Let ${\mathcal G}_{k}$ be the category of finite groups,
${\mathcal G}{\mathcal  A}lg_{k}$ the category of group
$k-$algebras. Let  ${\mathcal  M}_{k}$ the category of all square
matrices over $k.$ Objects of ${\mathcal  M}_{k}$ are natural
numbers and set of morphisms $Hom(n)$ is the  set of all square
matrices $k^{n \times n}.$ \\
{\em Grassmanian algebras.} \\
Let $GA(k)$ be the Grassmanian algebra over $k$ generated by the set $e_{1}, e_{2}, \dots .$
Each element $l \in GA(k)$ may be represented as a finite sum of an element $a \in k$ and elements
$e_{1} \wedge \dots \wedge e_{r}, r > 1,$ with coefficients in $k.$ The mapping
$l \mapsto f(l) = a$ defines an epimorphism $f:GA(k) \rightarrow k.$
 The monomial
$e_{1} \wedge \dots \wedge e_{r}$ is called {\em even} or {\em odd} depending on the
parity of $r.$ The monomial $1$ is considered to be even. Linear combinations of even (odd)
monomials with coefficients in $k$ form
the set $ GA _{e}$ of {\em even} elements (the set $ GA _{o}$ of {\em odd} elements) of the
algebra $GA(k).$ \\
{\em Hopf algebras}. \\
  Let $\cal H$ be an algebra with unit $e$ over field $k.$
Let $a, \; b \in {\cal H}, \; \mu(a,b) = ab$ the product in
$\cal H$, $\alpha \in k, \; p:k \rightarrow {\cal H},
\; p(\alpha) = \alpha e $ the unit, $\varepsilon: {\cal H}
\rightarrow k, \; \varepsilon(a) = 1, $
the counit, $\Delta(a) = a \otimes a$ the coproduct, $S: \; {\cal H}
\rightarrow {\cal H}$  the antipode, such that the axioms \\
1)$(1 \otimes \Delta) \circ \Delta = (\Delta \otimes 1) \circ \Delta$
(coassociativity); \\
2)$\mu \circ (1 \otimes S) \circ \Delta = \mu \circ
(S \otimes 1) \circ \Delta = p \circ \varepsilon$ (antipode). \\
Then the system $({\cal H}, \mu, p, \varepsilon, \Delta, S)$
is called the {\em Hopf algebra}.        \\
 Let ${\cal A}$ and ${\cal B}$ be
$C^*$-algebras. Let ${\cal K}$ be the $C^*$-algebra
of compact operators. If ${\cal A} \otimes {\cal K}$ is
isomorphic to ${\cal B} \otimes {\cal K}$ then ${\cal A}$
is called {\em Morita equivalent} to ${\cal B}$.

\subsubsection{$A_{\infty}-$algebras}

Strong homotopy algebras have been introduced by Stasheff. Strong homotopy
algebras, or shortly $A_{\infty}-$algebras have been  investigated in the context
of  algebras and superalgebras by Gugenheim, Stasheff, Penkava, Schwarz, Kontsevich,
Barannikov, Merkulov and others.

\subsection{SHEAVES AND MANIFOLDS}

In the section we follow to \cite{Sha88,GrH78,GeM:88,Le,Ma}. At first
collect some notions related to ringed spaces. \\
 The {\it ringed space} is the pair $(X,{\cal O}),$ where $X$ is a
topological space and $\cal O$ is a sheaf of rings on $X.$ For a
ringed space $(X,{\cal O}_{X})$ and an open $U \subset X$ the
restriction of the sheaf ${\cal O}_{X}$ on $U$ defines the ringed
space $(U,{\cal O}_{X|U}).$ \\ Let ${\cal O}_{X}$ be the sheaf of
smooth functions on a topological space $X.$ Then any smooth
manifold $X$ with the sheaf ${\cal O}_{X}$ is a ringed space
$(X,{\cal O}_{X}).$ \\

{\bf Example :}
   Let $X$ be a Hausdorff topological space and
${\cal O}_{X}$ a sheaf on $X.$ Let it satisfies conditions: (i)
${\cal O}_{X}$ is the sheaf of algebras over $\bf C$~; (ii) ${\cal
O}_{X}$ is a subsheaf of the sheaf of continuous complex valued
functions. Let $W$ be a domain in ${\bf C}^n$ and ${\cal O}_{an}$
the sheaf of analytical functions on $W.$ The
 ringed space $(X,{\cal O}_{X})$ is called the {\it complex analytical
manifold} if for any point $x \in X$ there exists a neighbourhood
$U \ni x$ such that $(U,{\cal O}_{X|U}) \simeq (W,{\cal O}_{an})$
(here $\simeq$ denotes the isomorphism of ringed spaces). \\
   Let $X$ be a projective algebraic variety over ${\bf C}$ and
${\cal C}oh_{X}$ the category of coherent sheaves on $X.$ \\

{\bf Example :} If $X$ is the algebraic curve then indecomposable
objects of ${\cal C}oh_{X}$ are skyscrapers and indecomposable
vector bundles. \\

\subsubsection{Supermanifolds}

Let $k^{n|m}$ be the {\em linear superspace} of all sequences
$(z_{1}, \ldots, z_{n}, \; \xi_{1}, \ldots , \xi_{m}),$
where $z_{i}$ and $\xi_{j}$ are  even and odd coordinates respectively.
 It is possible to represent coordinates  $z_{i}$ and $\xi_{j}$ by different
elements of  sets  $ GA _{e}$ and $ GA _{o}$ respectively.
In the case the mapping
$f((z_{1}, \ldots, z_{n}, \; \xi_{1}, \ldots , \xi_{m})) =
(f(z_{1}), \ldots, f(z_{n}))$ defines a morphism $f:k^{n|m} \rightarrow k^{n}.$
There are several possible ways of defining superspaces and contructing
supermanifolds and supervarieties. For instance one way of contructing
Riemann supervarieties is to use superspace $H_{1} = \{(z,\xi)
\in {\bf C}^{1|1} | Im f(z) > 0\}$ or superspace $H_{2} = \{(z,\xi_{1}, \xi_{2})
\in {\bf C}^{1|2} | Im f(z) > 0\}$ and corresponding super-Fuchsian groups.
As we consider mainly the complex case $k = {\bf C},$ recall the P. Deligne's
definition:  the {\em complex superspace}  is the space $X$ locally ringed
by a sheaf of commutative superalgebras ${\cal O}_{X} = {\cal O}_{0} \oplus {\cal O}_{1}$ such
that the pair $(X, {\cal O}_{0})$  is the usual complex space an $ {\cal O}_{1}$ is a
coherent sheaf of  ${\cal O}_{0}-$modules.

\subsubsection{Vector Bundles over Projective Algebraic Curves }

Let $X$ be a projective algebraic curve over algebraically
closed field $k$ and $g$ the genus of $X$.
Let ${\cal VB}(X)$ be the category of vector bundles over
$X$. Grothendieck have shown that for a rational curve every
vector bundle is a direct sum of line bundles. Atiyah have classified
vector bundles over elliptic curves. The main result is  \\
{\bf Theorem}~\cite{At}. Let $X$ be an elliptic curve, $A$ a
fixed base point on $X$. We may regard $X$ as an abelian
variety with $A$ as the zero element.  Let ${\cal E}(r,d)$ denote
the the set of equivalence classes of indecomposable vector
bundles over $X$ of dimension $r$ and degree $d$. Then each
${\cal E}(r,d)$ may be identified with $X$ in such a way that \\
$ det: {\cal E}(r,d) \rightarrow {\cal E}(1,d) $ corresponds to
$ H: X \rightarrow X, $ \\
where $ H(x) = hx = x + x + \cdots + x \; (h \:$ times$)$, and
$h = (r,d)$ is the highest common factor of $r$ and $d$. \\
Curve $X$ is called a {\it configuration} if its normalization
is a union of projective lines and all singular points of $X$
are simple nodes. For each configuration $X$ can assign a
non-oriented graph $\Delta(X)$, whose vertices are irreducible
components of $X$, edges are its singular and an edge is
incident to a vertex if the corresponding component contains the
singular point. Drozd and Greuel have proved: \\
{\bf Theorem}~\cite{DG}. 1. ${\cal VB}(X)$ contains finitely many
indecomposable objects up to shift and isomorphism if and only if
$X$ is a configuration and the graph $\Delta(X)$ is a simple chain
(possibly one point if $X = {\bf P}^1$). \\
2. ${\cal VB}(X)$ is tame, i.e. there exist at most one-parameter
families of indecomposable vector bundles over $X$, if and only if
either $X$ is a smooth elliptic curve or it is a configuration
and the graph $\Delta(X)$ is a simple cycle (possibly, one loop
if $X$ is a rational curve with only one simple node).   \\
3. Otherwise ${\cal VB}(X)$ is wild, i.e. for each finitely generated
$k-$algebra $\Lambda$ there exists a full embedding of the category
of finite dimensional $\Lambda-$modules into ${\cal VB}(X)$. \\

\subsection{Homological algbra of connections on manifolds and variation of Hodge structure}

The section follows ideas and results by Grothendieck, Griffiths, Manin, Katz, Deligne and others.
Let $S/k$ be the smooth scheme over field $k$, $U$ an element of
open covering of $S$, $ {\cal O}_S $ the structure sheaf on $S$,
$\Gamma(U,{\cal O}_{S})$ the sections of ${\cal O}_{S}$ on $U$.
Let $\Omega^{1}_{S/k}$ be the sheaf of germs of $1-$dimension
differentials, $\cal F$ a coherent sheaf on $S$. The {\em connection} on the sheaf
$\cal F$ is the sheaf homomorphism
$$ \nabla: {\cal F} \rightarrow \Omega^{1}_{S/k} \otimes {\cal F},$$
such that, if $ f \in \Gamma(U,{\cal O}_{S}), \;
g \in \Gamma(U,{\cal F})$ then
$$ \nabla(fg) = f\nabla(g) + df \otimes g.$$
  There is the dual definition. Let ${\cal F}$ be the locally
free sheaf, $\Theta^{1}_{S/k}$ the dual to sheaf $\Omega^{1}_{S/k}$,
$\partial \in \Gamma(U,\Theta^{1}_{S/k})$.
The {\em connection} is the homomorphism
$$ \rho: \Theta^{1}_{S/k} \rightarrow
End_{{\cal O}_S}({\cal F},{\cal F}), $$
$$\rho(\partial)(fg) = \partial(f)g + f\rho(\partial). $$

\subsubsection {Integration of connection}

Let $\Omega^{i}_{S/k}$ be the sheaf of germs of $i-$differentials,
$$ \nabla^{i}(\alpha \otimes f) = d\alpha \otimes f + (-1)^{i}\alpha
\wedge \nabla(f). $$
   Then $ \nabla, \; \nabla^{i}$ define the sequence of homomorphisms:
\begin{equation}
 \label{CC}
 {\cal F} \rightarrow \Omega^{1}_{S/k} \otimes {\cal F}
 \rightarrow \Omega^{2}_{S/k} \otimes {\cal F} \rightarrow \cdots     ,.
\end{equation}
The map
$K = \nabla \circ \nabla^{1}: {\cal F}  \rightarrow \Omega^{2}_{S/k} \otimes {\cal F}$ is called
the curvature of the connection $\nabla.$

The {\it cochain complex}
\begin{equation}
 (K^{\bullet},d) = \{K^{0}
\stackrel{d}\rightarrow K^{1} \stackrel{d}\rightarrow K^{2}
\stackrel{d}\rightarrow \cdots \}
\end{equation}
is the sequence
of abelian groups and differentials $d: K^{p} \rightarrow K^{p+1}$
with the condition $d \circ d = 0.$
  A connection is {\em integrable} if (\ref{CC}) is a complex.\\

{\bf Proposition.}
 The statements a),b), c) are equivalent: \\
a)  the connection $\nabla$ is  integrable;\\
b)  $K = \nabla \circ \nabla^{1} = 0;$\\
c)  $ \rho $ is the Lie-algebra homomorphism of sheaves of Lie algebras.

\subsection{Topological and Homotopical categories}

Let $X$ be a compact smooth manifold with boundary $\partial X.$
Let $\partial X = \partial X_{0} \bigcup \partial X_{1}$ be a partition
such that $\partial X_{0}$ and $\partial X_{1} $ are integer components of
$\partial X. $ In the case the manifold $X$ is called  the {\em cobordism}
with the source $\partial X_{0}$ and target $\partial X_{1}.$ The case
$\partial X_{0} = \emptyset$ and $\partial X_{1} = \emptyset$ means
that $X$ is closed.
Let $MX$ be the class of $n-$dimensional compact smooth manifolds with boundary
and $OX$  the class of the connected components of the boundaries. From these
data form the category ${\bf CB}_{n}$ whose objects are connected components of
the boundaries of $n-$dimensional manifolds. For any two boundaries
$\partial Y_{0}$ and $\partial Y_{1} $ a morphism between them is a manifold
$Y \in MX$ such that $\partial Y = \partial Y_{0} \bigcup \partial Y_{1}.$
If we regard two morphisms in ${\bf CB}_{n}$ as equivalent if they are
homotopic then we can  form the quotient category ${\bf hCB}_{n}.$
 Respectively for a given compact smooth manifold $X$  with boundary it is
possible to form a category ${\bf CB}_{X}$ whose morphisms are $n-$dimensional
submanifolds of $X$ and whose objects are connected components of boundaries
of such submanifolds.  The category ${\bf hCB}_{X}$ have the same objects as
${\bf CB}_{X}$ and morphisms of ${\bf hCB}_{X}$ are homotopy equivalent
classes of morphisms of ${\bf CB}_{X}.$  \\

Recall now the relevant properties of complexes, derived
categories, cohomologies and quasimorphisms referring
to~\cite{GrH78,GeM:88} for details and indication of proofs. \\

 Let ${\cal A}$ be an abelian
category, ${\cal K}({\cal A})$ the category of complexes over
${\cal A}$. Furthermore, there are various full subcategories of
${\cal K}({\cal A})$ whose respective objects are the complexes
which are bounded below, bounded above, bounded in both sides. The
notions of homotopy morphism, homotopical category, triangulated
category are described in \cite{GeM:88}. The bounded derived
category ${\mathcal D}^{b}(X)$ of coherent sheaves on $X$ has the
structure of a triangulated category \cite{GeM:88}.
\subsubsection{The Fourier-Mukai transform}
Let $A$ be an abelian variety and ${\hat A}$ the dual abelian
variety which is by definition a moduli space of line bundles of
degree zero on $A.$
The {\em Poincar{\'e}
bundle} ${\mathcal P}$ is a line bundle of degree zero on the
product $A \times {\hat A}$, defined in such a way that for all $
a \in {\hat A}$ the restriction of ${\mathcal P}$ on $A \times
\{a\}$ is isomorphic to the line bundle corresponding to the point
$a \in {\hat A}.$ This line bundle is also called {\em the
universal bundle}. Let
\[
{\pi}_{A}: \; A \times {\hat A} \rightarrow A,
\]
\[
{\pi}_{\hat A}: \; {\hat A} \times  A \rightarrow {\hat A},
\]
and ${\mathcal C}_{A}$ be the category of $O_{A}-$modules over $A,
\; M \in Ob \; {\mathcal C}_{A},$
\[
  {\hat S}(M) = {\pi}_{{\hat A},*}({\mathcal P} \otimes
  {\pi}_{A}^{*} M).
\]
Then, by definition, {\em the Fourier-Mukai transform} ${\mathcal
F}{\mathcal M}$ is the derived functor $R{\hat S}$ of the functor
${\hat S}.$  Let ${\mathcal D}(A), \; {\mathcal D}({\hat A})$ be
bounded derived categories of coherent sheaves on $A$ and ${\hat
A}$ respectively.\\
{\bf Theorem}(Mukai.)
 The derived functor ${\mathcal F}{\mathcal M} = R{\hat S}$
induces an equivalence of categories between two derived
categories ${\mathcal D}(A)$ and ${\mathcal D}({\hat A}).$

\subsection{PROPs, Operads, $A_{\infty}-$categories and $A_{\infty}-$functors}
PROPs was introduced  by Mac Lane and Adams.
Their topological versions was defined by Bordman and explored by
Bordman and Vogt~\cite{BV:IA}.
An interval PROP $P$ is a category with objects $0, 1, 2, ...$
 satisfying the conditions from\cite{BV:IA} and is defined
 on interval manifolds. It has representation by generalized trees
 that are defined and take values on interval manifolds.  \\
 Interval PROPs have $A_\infty$-category properties.  \\
Operads was introduced  by J. May~\cite{M:GI}.
Here we give a short description of interval operad.
 The space of continuous interval
functions of $j$ variables forms the topological space $I{\cal C}(j).$
Its points are operations $I^{j}{\bf R} \to I{\bf R}$ of arity  $j.$
$I{\cal C}(0)$ is a single point $*.$ The class of interval spaces
$I^{n}{\bf R}, \; n \ge 0$ forms the category~\cite{Gl:ICC}. We will consider
the spaces with base points and denote the category of those spaces by
$I{\cal U}.$  Let $X \in I{\cal U}$ and for $k \ge 0$ let $I{\cal E}(k)$
be the space of maps $M(X^{k},X).$ There is the action (by permuting the
inputs) of the symmetric group $S_k$ on $I{\cal E}(k).$ The identity
element $1 \in I{\cal E}(1)$ is the identity map of $X.$
 In the above mentioned conventions let $k \ge 0$ and $j_{1}, \ldots ,
j_{k} \ge 0$ be integers. Let for each choice of $k$ and  $j_{1}, \ldots ,
j_{k}$  there is a map
\begin{eqnarray*}
\gamma: I{\cal E}(k) \times I{\cal E}(j_{1}) \ldots \times I{\cal E}(j_{k})
\to I{\cal E}(j_{1} + \ldots + j_{k})
\end{eqnarray*}
given by multivariable composition. If maps $\gamma$ satisfy associativity,
equivalence and unital properties then $I{\cal E}$ is the endomorphism
interval operad $I{\cal E}_X$ of $X.$
Here all categories will be assumed to
be {\em linear} over $k$ and small. A {\em non-unital} $A_{\infty}-${\em category} ${\bf A}$
consists of a set of objects
$Ob {\bf A}$, a graded vector space $hom_{\bf A}(X_{i}, X_{j})$ for any pair of objects, and composition
maps of every order $d \ge 1$,\\
$\mu_{\bf A}^{d}: hom_{\bf A}(X_{d-1}, X_{d}) \otimes \ldots \otimes hom_{\bf A}(X_{0}, X_{1})
\rightarrow \\
hom_{\bf A}(X_{0}, X_{d})[2-d]. $\\
where $[s]$ means shifting the grading of a vector space down by $s \in {\bf Z}$. The
maps must satisfy the (quadratic) $A_{\infty}-$associativity equations~\cite{BV:IA}.


\subsection{Integrability and foliations}

Integrability is connected with foliations.
A simplest example of foliation is a trivial $k$ dimensional
foliation or a trivial codimension $n-k$ foliation of Euclidean
space ${\bf R}^{n}={\bf R}^{k} \times {\bf R}^{n-k}, 0 \le k \le
n:$
$$ {\bf R}^{n} = \bigcup_{(x_{k+1},\ldots ,x_{n})}{\bf R}^{k}
\times (x_{k+1},\ldots,x_{n}), $$
that is, $ {\bf R}^{n}$ decomposes into a union of
$ {\bf R}^{k} \times (x_{k+1}, \ldots ,x_{n}) $ 's each of
which is $ C^{\infty}$ diffeomorphic to $ {\bf R}^{k}.$
$ {\bf R}^{k} \times (x_{k+1}, \ldots ,x_{n}) $ is called
a leaf of the foliation. \\
Still one example can be obtained from consideration of
nonsingular vector fields on the torus.
More generally,
a nonsingular flow on a manifold  corresponds to a foliation
on the manifold by one dimensional leaves where the leaves
are provided with Riemannian metrics and directed.
 
Let $(M, {\cal F})$ be a foliated
manifolds, $(M / {\cal F})$ its leaf space, and $G = G(M, {\cal F})$ the holonomy
groupoid of  $(M / {\cal F})$ with $s, r$ the source and target maps.
 Recall follow to~\cite{Mo:OA} some fact about
foliation $C^{*}-$algebras and holonomy groupoids. Let $C_{c}(G)$ denote the space
of continuous functions on $G$ with compact support. It is possible define
a convolution product on $C_{c}(G)$. Given $\gamma \in G,\; \varphi \in C_{c}(G),$
 the involution $*$ on $C_{c}(G)$ is defined as
$$ \varphi^{*}(\gamma) = \overline{\varphi}(\gamma^{-1}).$$
Denote by $x$ any element $M.$
 Let $L_{x}$ be the leaf that contains $x$ and
${\tilde L}_{x} = \{\gamma \in G | s(\gamma) = x \}.$
The target map $r$ yields the covering progection $r: {\tilde L}_{x} \rightarrow L_{x}.$
Denote by $d\mu = \{d\mu_{x}\}_{x \in M}$ the family of measures on
$\{{\tilde L}_{x} \}_{x \in M}.$
Let ${\cal H}_{x}$ be the Hilbert space of $L^{2}-$functions on
${\tilde L}_{x}, \xi \in {\cal H}_{x}, \alpha, \beta \in {\tilde L}_{x}.$
A $*-$representation ${\pi}_{x}$ on ${\cal H}_{x}$ is defined by
$$ [{\pi}_{x}(\varphi)\xi](\alpha) =
\int\limits_{{\tilde L}_{x}} \varphi(\alpha {\beta}^{-1}) \xi(\beta) {d\mu}_{x}({\beta}).$$
Let $\parallel \varphi \parallel$ be the operator norm of ${\pi}_{x}(\varphi)$ on
${\cal H}_{x}.$

For a given foliated manifold $(M, {\cal F})$ the foliated $C^{*}-$algebra
$C^{*}(M, {\cal F})$ is the  $C^{*}-$complation of $C_{c}(G)$ with respect to
the $*-$norm.

\subsection{To TQFT}
 Let $M$ be an $n-$dimensional manifold and ${\bf hCB}_{n}$ the quotient category of  ${\bf CB}_{n}.$
A topological quantum field theory (TQFT)
 on $M$ is a symmetric monoidal functor from ${\bf hCB}_{n}$ to the category of vector spaces.
A TQFT on $n-$dimensional manifolds is then a functor from ${\bf hCB}_{n}$ to the category of
vector spaces, which takes disjoint unions of bordisms to the tensor product of corresponding vector
spaces. \\
 Let $AM$ be the associative algebra of functions on $M$ defined by the product
of functions and $C^{*}(AM)$ the Hochschild complex. A $p-$cochain $C$
on $AM$ is a $p-$linear map of $AM^{p}$ into $AM.$ A $1-$cocycle is a
derivation of the algebra, that is a vector field. The Hochschild coboundary
 of $C$ is the $(p-1)-$cochain $\partial C$ and $\partial^2 = 0.$  The complex
$C^{*}(AM)$ has  the structure of homotopic Gerstenhaber algebra. The proof
is found in ~\cite{VG:HO}. \\
Let $H^{p}(AM;AM)$ be the $p-$th Hochschild cohomology space for the considered
differentia complex. By Vey's result~\cite{V:P} $H^{p}(AM;AM)$ is isomorphic to
 the space of the antisymmetric contravariant $p-$tensors of $M.$



\end{document}